# Coulomb blockade in an atomically thin quantum dot coupled to a tunable Fermi reservoir

Mauro Brotons-Gisbert[1*], Artur Branny[1†], Santosh Kumar[1⊥], Raphaël Picard[1], Raphaël Proux[1], Mason Gray[2], Kenneth S. Burch[2], Kenji Watanabe[3], Takashi Taniguchi[3], Brian D. Gerardot[1*]

[1]*Institute for Photonics and Quantum Sciences, SUPA, Heriot-Watt University, Edinburgh EH14 4AS, UK.*
[2]*Boston College Department of Physics, 140 Commonwealth Avenue, Chestnut Hill, MA 02467, USA.*
[3]*National Institute for Materials Science, 1-1 Namiki, Tsukuba, 305-0044, Japan.*
[†]*Current address: Department of Applied Physics, Royal Institute of Technology, Stockholm 106 91, Sweden*
[⊥]*Current address: Indian Institute of Technology Goa GEC Campus, Ponda-403401, Goa, India*
[*]*e-mail address: M.Brotons_i_Gisbert@hw.ac.uk or B.D.Gerardot@hw.ac.uk*

**Gate-tunable quantum-mechanical tunnelling of particles between a quantum confined state and a nearby Fermi reservoir of delocalized states has underpinned many advances in spintronics and solid-state quantum optics. The prototypical example is a semiconductor quantum dot separated from a gated contact by a tunnel barrier. This enables Coulomb blockade, the phenomenon whereby electrons or holes can be loaded one-by-one into a quantum dot[1,2]. Depending on the tunnel-coupling strength[3,4], this capability facilitates single spin quantum bits[1,2,5] or coherent many-body interactions between the confined spin and the Fermi reservoir[6,7]. Van der Waals (vdW) heterostructures, in which a wide range of unique atomic layers can easily be combined, offer novel prospects to engineer coherent quantum confined spins[8,9], tunnel barriers down to the atomic limit[10] or a Fermi reservoir beyond the conventional flat density of states[11]. However, gate-control of vdW nanostructures[12–16] at the single particle level is needed to unlock their potential. Here we report Coulomb blockade in a vdW heterostructure consisting of a transition metal dichalcogenide quantum dot coupled to a graphene contact through an atomically thin hexagonal boron nitride (hBN) tunnel barrier. Thanks to a tunable Fermi reservoir, we can deterministically load either a single electron or a single hole into the quantum dot. We observe hybrid excitons, composed of localized quantum dot states and delocalized continuum states, arising from ultra-strong spin-conserving tunnel coupling through the atomically thin tunnel barrier. Probing the charged excitons in applied magnetic fields, we observe large gyromagnetic ratios (~8). Our results establish a foundation for engineering next-generation devices to investigate either novel regimes of Kondo physics or isolated quantum bits in a vdW heterostructure platform.**

Our device, shown schematically in Fig. 1a, consists of a quantum dot in monolayer WSe$_2$ separated from a Fermi reservoir in few-layer graphene by a monolayer hBN tunnel barrier. The WSe$_2$ is fully encapsulated on the bottom side by hBN (see Supplementary Fig. 1). This heterostructure was mechanically stacked in an inert environment on an insulating SiO$_2$ layer on a n-doped Si substrate (back gate). Gate-tuning is achieved by applying a bias $V_g$ between the graphene electrode and the grounded back gate. Confocal photoluminescence imaging of the sample at a temperature of 3.8 K and $V_g = 0$ V reveals a few localized spots with higher photoluminescence intensity than the homogeneous background photoluminescence (Fig. 1b). The localized bright spots show discrete spectrally narrow peaks arising from WSe$_2$ quantum emitters[14-16] that are spectrally and spatially isolated due to local strain[17–20]. Here, local strain is provided from a wrinkle in the bottom hBN layer (Supplementary Fig. 1). Figure 1c shows $V_g$-dependent photoluminescence spectra measured at the brightest spot in Fig. 1b. Near $V_g = 0$ V, we observe three spectrally narrow lines corresponding to the neutral excitons (X$^0$) of three different optically active quantum dots (labelled A to C). Notably, the neutral exciton energy of dots A, B and C is independent of the vertical electric field across the device, demonstrating minimal quantum confined Stark effect for these WSe$_2$ quantum dots, in contrast to previous reports for WSe$_2$ quantum dots[21] but similar to two-dimensional excitons in TMDs[22]. At $V_g \approx -7$ V (dot A) and $V_g \approx -13$ V (dots B and C), the emission energy changes abruptly as a second electron overcomes the electron–electron Coulomb energy ($U_{ee}$) and tunnels into the quantum dot (schematically represented in the top part of Fig. 1d), creating negatively charged excitons (X$^{1-}$) with binding energies of ~25 meV (dot A) and ~23 meV (dots B and C). Additionally, for dot B we observe a spectral jump at $V_g > 10$ V as a second hole overcomes the hole–hole Coulomb energy ($U_{hh}$) and tunnels into the quantum dot (schematically represented in the bottom of Fig. 1d) to create the positively charged exciton (X$^{1+}$) with 7 meV binding energy. These results demonstrate an unambiguous Coulomb blockade at the single particle level and the unique ability to tune the Fermi reservoir from n-type to p-type in a vdW heterostructure.

To elucidate the nature of WSe$_2$ quantum dots, their strong tunnel coupling to the tunable Fermi reservoir and the consequences on the excitonic states, we focus in detail on dot B, which exhibits both the X$^{1-}$ and X$^{1+}$ at reasonably modest $V_g$. Figure 2a shows $V_g$-dependent high-resolution photoluminescence spectra. At $V_g = 0$ V, the X$^0$ exhibits a doublet split by ∼830 μeV with orthogonally linear polarized emission (see Supplementary Fig. 4). Conversely, for X$^{1-}$ and X$^{1+}$ (at $V_g < -9$ V and $V_g > 16$ V, respectively) we observe a single spectral line, in contrast to a previous report[23]. The inset in Fig. 2a shows a diagram representing the exciton states. For X$^0$, the doublet is a fine-structure splitting (FSS) arising due to electron–hole exchange interaction energy ($\Delta_{FSS}$), commonly observed for neutral excitons in InAs/GaAs[2,24] or WSe$_2$ (refs. [14–17]) quantum dots. On adding a second electron or hole to the neutral exciton, the minority particle interacts with a spin singlet and the exchange interaction vanishes. Notably, the exchange interaction only disappears in charged excitons for quantum dots in the strong confinement regime[24], identifying the nature of the monolayer WSe$_2$ quantum dots.

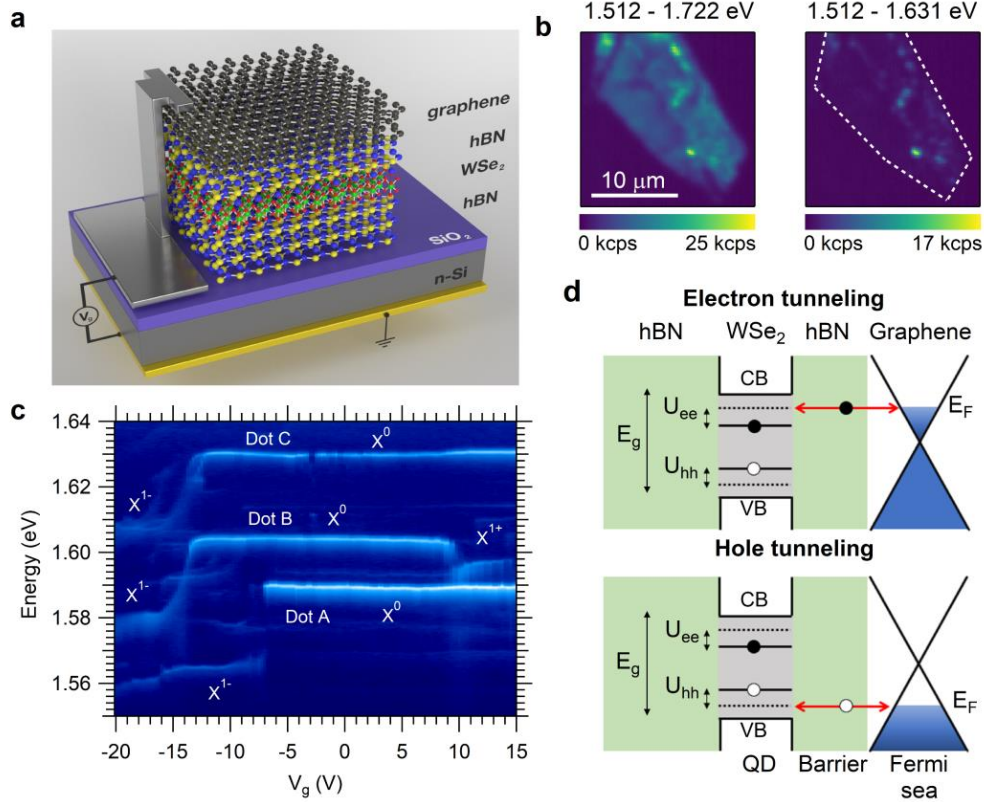

**FIG. 1. Coulomb blockade in a vdW heterostructure device. a**, Sketch of the vdW charge-tunable quantum dot device. **b**, Colour-coded spatial map of the integrated photoluminescence signal of the WSe$_2$ in the spectral range of 1.512–1.722 eV (left) and 1.512–1.631 eV (right), highlighting the full flake and the single quantum emitters, respectively. The localized bright spots in the right panel correspond to quantum dots. The white dashed line indicates the edges of the WSe$_2$ monolayer region. kcps, kilo counts per second. Scale bar, 10 μm. **c**, Voltage-dependent low-resolution photoluminescence of three quantum dots at the brightest spot in **b** showing neutral (X$^0$), negatively charged (X$^{1-}$) and positively charged (X$^{1+}$) exciton species and Coulomb blockade. The unlabelled spectral features with weak peak intensity in **c** originate from distinct quantum dots unrelated to dots A, B and C (see Supplementary Information). **d**, Schematic representation of the electron (filled circles) and hole (open circles) tunnelling through the hBN barrier from the Fermi reservoir in the graphene to the quantum dots in WSe$_2$. CB, VB and $E_g$ represent the conduction band, valence band and the energy bandgap of the WSe$_2$ quantum dots, respectively. $U_{ee}$ ($U_{hh}$) represents the electron–electron (hole–hole) Coulomb interaction energy. QD, quantum dot.

Numerous signatures of the strong coupling regime between both electrons and holes in the quantum dot and the graphene Fermi reservoir can be found in the $V_g$-dependent photoluminescence spectra. First, the charging steps are abrupt in voltage, indicating that the tunnelling rate is much faster than exciton recombination rate (∼1–10 ns for WSe$_2$ quantum dots[14–17]). Second, the energies of the X$^0$ doublet peaks redshift near the edges of the plateau, indicating hybridization of energy levels in the quantum dot and Fermi reservoir. Third, the continuous smooth transition in energy from the X$^{1-}$ to the X$^0$ states signifies the existence of a hybrid exciton (X$_H$) arising due to strong mixing between the wavefunctions of the discrete quantum dot-states and the continuum of states of the Fermi reservoir[7,25,26]. Fourth, at the crossover point from $X^0$ to $X^{1+}$ the $X^{1+}$ energetically bends and joins another transition line labelled $X_{f+}$ that arises due to hybridization of the hole with a continuum of states in the Fermi reservoir[27]. Finally, Lorentzian line shapes are observed for the $X^0$ states far from the hybridization regime, whereas the hybrid and charged excitons exhibit broad and highly asymmetric photoluminescence line shapes with prominent low-energy tails (see Supplementary Fig. 5). These low-energy tails are a consequence of the Anderson orthogonality catastrophe: an energy shakeup process experienced by the electrons in the conduction band (holes in the valence band) when the hybrid excitons recombine and change the electron (hole) level in the quantum dot due to the sudden removal of the intra-quantum-dot Coulomb attraction with the localized hole (electron)[7,28,29]. We apply the Anderson impurity model to extract the tunnel-coupling strength in the device and explain the evolution of each exciton state as a function of $V_g$. We achieve quantitative agreement with this evolution under the assumption of that the Fermi reservoir is reduced to zero bandwidth at $E_F$ (see Supplementary Equations (1)–(15)). To properly account for the asymmetric photoluminescence line shapes and the continuous transition from $X_{f+}$ to $X^{1+}$, a more sophisticated calculation involving the density of states in the Fermi reservoir is required[7,27,29]. This calculation and the impact of the Fermi reservoir density of states on the many-body interactions are left for future investigation.

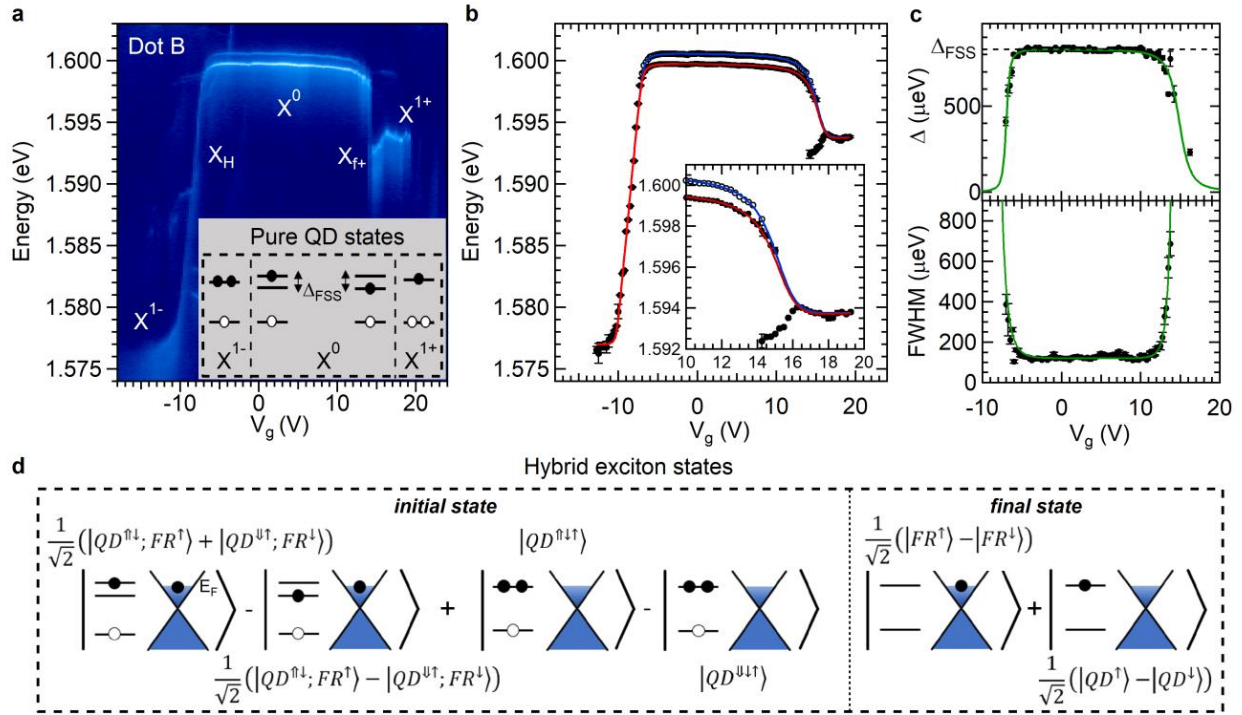

**FIG. 2. Strong tunnel coupling between a quantum dot and a tunable Fermi reservoir in a vdW heterostructure. a**, Voltage-dependent high-resolution photoluminescence of dot B. The inset shows a diagram that represents the energy levels of the electrons and the occupation of holes for the pure $X^{1-}$, $X^0$ and $X^{1+}$ exciton states of the quantum dot. $\Delta_{FSS}$ represents the energy splitting of the fine-structure split $X^0$ state. **b**, Voltage-dependent evolution of the emission energy for the quantum dot shown in **a**. The solid lines represent fits of the experimental data using the zero-bandwidth model. Details of the theoretical model can be found in Supplementary Equations (1)–(15). The inset displays a zoom of the transition from the $X^0$ to the $X^{1+}$ exciton state. **c**, Upper panel: experimental (filled circles) and calculated (green line) evolution of the energy splitting ($\Delta$) of the $X^0$ doublet as a function of $V_g$. Lower panel: experimental (filled circles) and calculated (green line) voltage-dependent evolution of the photoluminescence linewidth for the low-energy peak of $X^0$. **d**, Schematic representation of the initial and final states of the $X^{1-}$ to $X^0$ hybrid exciton. The single (double) arrows represent the spin orientation of a single electron (hole).

Figure 2d diagrams the initial and final states for the hybridization of the $X^0$ and $X^{1-}$ states. Filled (open) circles represent electrons (holes), whereas the superscript single (double) arrows account for the possible orientations of the electron (hole) spin. The initial state, composed of two electrons and one hole, is a superposition of different excitonic states: (1) two states that correspond to the quantum dot containing $X^0$ and an additional electron in the Fermi reservoir, in which the exchange interaction energetically splits the $X^0$ into two states by $\Delta_{FSS}$ $((|QD^{\Uparrow\downarrow}, FR^\uparrow\rangle \pm |QD^{\Downarrow\uparrow}, FR^\downarrow\rangle)/\sqrt{2})$; and (2) two states that correspond to the excitonic configurations in which the quantum dot has two electrons with opposite spin orientations and a hole ($|QD^{\Uparrow\uparrow\downarrow}\rangle$ and $|QD^{\Downarrow\uparrow\downarrow}\rangle$), giving rise to $X^{1-}$ (with no exchange interaction). After photon emission from the initial state, the final state contains only one electron which is in a superposition of two states: (1) a state corresponding to a linear combination of the spin orientation states of the remaining electron in the Fermi reservoir ($|FR^\uparrow\rangle$ and $|FR^\downarrow\rangle$), and (2) a state that is a superposition of the spin orientations for the remaining electron in the quantum dot ($|QD^\uparrow\rangle$ and $|QD^\downarrow\rangle$). The individual states in both the initial and final state are coupled by a spin-conserving tunnel interaction (see Supplementary Equations (1–15)).

The solid lines in Fig. 2b represent fits of the experimental data to our model, capturing the photoluminescence evolution for both hybrid and bare quantum dot excitons states over the full range of $V_g$ values with high accuracy. The fits reveal an identical lever arm (ratio of device thickness to tunnel barrier thickness) of $\lambda = 145 \pm 10$ for electrons and holes in our device, from which an hBN tunnel barrier thickness of 0.5 ± 0.2 nm (corresponding to one or two monolayers) is obtained. The tunnel coupling energies for both electrons ($V_{tun}^e$) and holes ($V_{tun}^h$) can be directly extracted from the fits; we find $V_{tun}^e = 1.2 \pm 0.2$ and $V_{tun}^h = 3.5 \pm 0.3$ meV. These values are one order of magnitude larger than has been possible in traditional III–V semiconductor devices[7,27]. We attribute the ultra-strong tunnel coupling to the reduction of the tunnel barrier thickness to the atomic layer limit. The ratio $V_{tun}^h/V_{tun}^e = 2.9 \pm 0.5$ is a consequence of the band alignment resulting from the monolayer $WSe_2$/hBN/graphene heterostructure, which leads to notably lower tunnel barrier heights for holes than for electrons[30]. Using the Wentzel–Kramer–Brillouin approximation for a rectangular tunnelling barrier (see Supplementary Equation (20)), $V_{tun}^h/V_{tun}^e \sim 3.1 \pm 0.4$ can be estimated, in agreement with the experimental result. This approximation can also be employed to explain the tunnel-induced broadening of the photoluminescence linewidth as a function of $V_g$ (see Supplementary Equation (21)), as shown in the bottom panel of Fig. 2c for dot B.

The calculated results in Fig. 2b accurately capture the strong tunnel-induced redshift of the $X^0$ states at each edge of the $X^0$ plateau. Different energy shifts are observed for each peak of the doublet, such that the energy splitting of the fine-structure split $X^0$ excitonic states ($\Delta$) changes as a function of $V_g$. The top panel of Fig. 2d shows the experimental and calculated evolution of $\Delta$ along the $X^0$ plateau extracted from the data shown in Fig. 2b. In the $V_g$ range corresponding to the pure $X^0$ quantum dot state $\Delta = \Delta_{FSS}$. However, a fast decrease of $\Delta$ is observed near the tunnelling transitions (left and right edges of the $X^0$ plateau), a consequence of the energy difference that exists between the initial and final hybridized states associated to each of the fine-structure split $X^0$ states. This is a striking consequence of a large $\Delta_{FSS}$ and a strong tunnel coupling: the different eigenstates of $X^0$ couple differently to the Fermi sea. The zero-bandwidth model also predicts the existence of the $X_{f+}$ feature in the set of solutions (see Supplementary Equation (14)), although it introduces an artificial splitting between this solution and the experimental result (see Supplementary Fig. 6). Introducing a finite bandwidth for the Fermi reservoir can more accurately model the $X_{f+}$ feature[27] (see Supplementary Fig. 8).

To further investigate the hybridization between the $X^0$ and $X^{1-}$ and between the $X^0$ and $X^{1+}$ exciton states, we apply a magnetic field $\mathbf{B}_z$ along the direction perpendicular to heterostructure interfaces (Faraday geometry). Figure 3a shows the result for $\mathbf{B}_z$ =5.3 T; the $X^{1-}$, $X^0$ and $X^{1+}$ quantum dot states Zeeman split due to the application of $\mathbf{B}_z$. We adapt our model to explore theoretically the hybridization between exciton states under $\mathbf{B}_z$ (see Supplementary Equations (16–19)). Figure 3b shows the $V_g$-dependent evolution of the energies for the high- (open circles) and low-energy (filled circles) emission peaks of dot B. The zero-bandwidth model quantitatively reproduces the measured evolution of the photoluminescence energy for ranges of $V_g$ values corresponding to both hybrid and pure quantum dot states. In contrast to the results for $\mathbf{B}_z = 0$ T, the tunnel-induced bending observed at the edges of the $X^0$ plateau is very similar for both of the fine-structure split $X^0$ states (Fig. 3a,b) and $\varDelta$ remains constant. This change in behaviour is a consequence of the reduction in the energy difference between the initial and final hybridized states due to the considerable Zeeman splitting. Notably, these results can only be modelled with spin-conserving tunnelling.

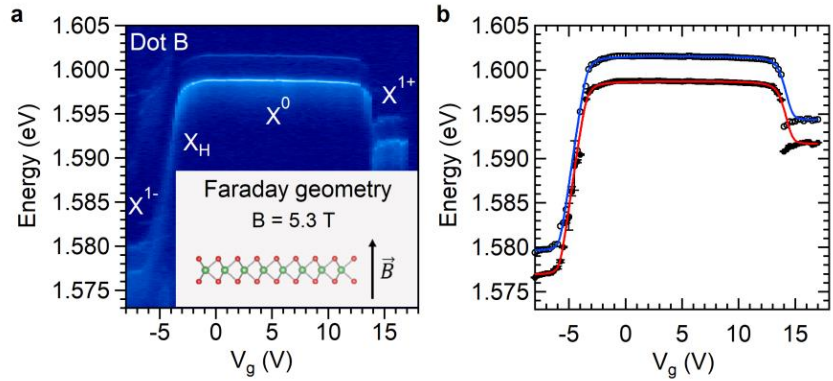

**Fig. 3. Strong tunnel coupling between a quantum dot and a tunable Fermi reservoir at high magnetic field. a**, Voltage-dependent high-resolution photoluminescence of dot B for $\mathbf{B}_z = 5.3$ T. The inset shows a schematic of the Faraday geometry. **b**, Voltage-dependent evolution of the high-energy (open circles) and low-energy (filled circles) emission peaks extracted from fits of the experimental data shown in **a**. The solid lines represent fits of the experimental data to the zero-bandwidth model described in the Supplementary Equations (16–19).

The ability to deterministically load either an electron or hole into the quantum dot allows us to magneto-optically probe the $X^{1-}$ and $X^{1+}$, respectively, in a $WSe_2$ quantum dot. Figure 4a shows photoluminescence spectra of the $X^{1-}$ (left panels), $X^0$ (central panels) and $X^{1+}$ (right panels) exciton states of dot B for varying $\mathbf{B}_z$ values, revealing a clear Zeeman splitting for each state. Figure 4b,c shows the magnetic-field dependence of the energy splitting measured for $X^0$, $X^{1-}$ and $X^{1+}$ of dot B and for $X^0$ and $X^{1-}$ of dot A, respectively. The results reveal that the charged excitons exhibit g factors of ~8.7 (dot B) and ~7.7 (dot A), mimicking the behaviour of the corresponding neutral excitons.

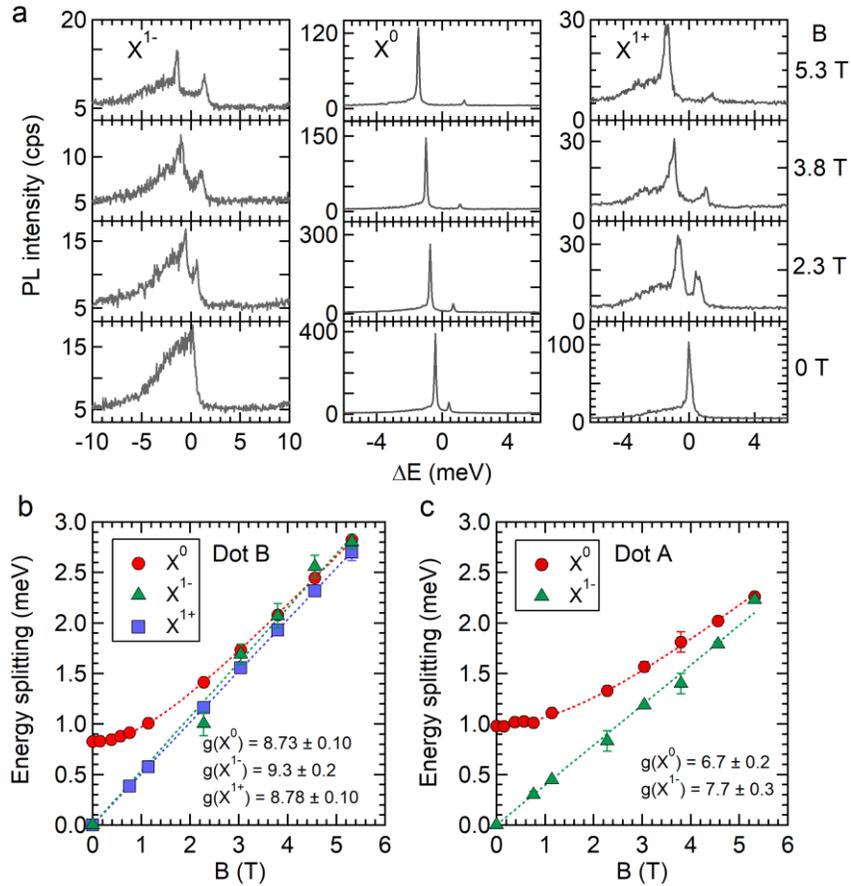

**Fig. 4. Magneto-optics of neutral and charged excitons in $WSe_2$ quantum dots. a**, Photoluminescence spectra of the $X^{1-}$ (left panels), $X^0$ (central panels) and $X^{1+}$ (right panels) exciton states of dot B under different applied magnetic fields in Faraday geometry. **b,c**, Magnetic-field dependence of the energy splitting measured for $X^0$, $X^{1-}$ and $X^{1+}$ of dot B (**b**) and for $X^0$ and $X^{1-}$ of dot A (**c**), as obtained from fits of the experimental data shown in **a**. Dashed lines represent fits of the experimental data to Supplementary Equations (22) ($X^0$) and (23) ($X^{1-}$ and $X^{1+}$). The fits reveal g values of 8.73 ± 0.10, 9.3 ± 0.2 and 8.78 ± 0.10 for the $X^0$, $X^{1-}$ and $X^{1+}$ exciton states of dot B, respectively, and g values of 6.7 ± 0.2 and 7.7 ± 0.3 for the $X^0$ and $X^{1-}$ of dot A, respectively.

Via a Coulomb blockade, we have demonstrated the ability to deterministically load a single electron or single hole in a vdW heterostructure quantum device. This is achieved with gate-tunable tunnel coupling between an optically active $WSe_2$ quantum dot and a tunable Fermi reservoir in few-layer graphene. Due to an atomically thin tunnel barrier, we obtain ultra-strong and spin-conserving tunnel coupling (roughly one order of magnitude stronger than in conventional III–V quantum devices) between the quantum dot and Fermi reservoir, leading to the observation of hybrid excitons that can be controlled by the gate voltage. Magneto-optical characterization of the charged excitons reveals large gyromagnetic ratios, indicating that both spin and valley degrees of freedom play an important role for single spins in $WSe_2$ quantum dots. These results confirm the potential of vdW heterostructures as a new platform for engineering quantum devices. On the one hand, with vdW heterostructures in the strong tunnel-coupling regime, the quantum confined states can be coupled to a tailored or tunable Fermi reservoir. This can enable high-fidelity electrical injection of polarized spins from a nearby ferromagnet or investigation of Kondo-phenomena beyond metallic-like Kondo screening. On the other hand, these results position vdW heterostructures as an intriguing platform to engineer a coherent spin-photon interface. In charge-tunable devices with larger tunnel barriers, a quantum dot can be isolated from its mesoscopic environment. Resonant excitation techniques[1,2,5,31] can then be used to probe and manipulate the valley and spin degrees of freedom and investigate their suitability as coherent quantum bits of information.

## ACKNOWLEDGEMENTS

**Funding:** This work is supported by the EPSRC (grant numbers EP/L015110/1, EP/P029892/1, and EP/M013472/1) and the ERC (numbers 307392 and 725920). Growth of hexagonal boron nitride crystals by KW and TT was supported by the Elemental Strategy Initiative conducted by the MEXT, Japan and the CREST (JPMJCR15F3), JST. Device fabrication by MG and KSB was made possible with support the National Science Foundation, Award No. DMR-1709987. BDG is supported by a Wolfson Merit Award from the Royal Society and a Chair in Emerging Technology from the Royal Academy of Engineering.
**Author contributions:** BDG conceived and supervised the project. AB fabricated the samples, assisted by SK, R. Picard, MG, and KSB. KW and TT supplied the boron nitride crystals. MBG and AB performed the experiments, assisted by SK and R. Proux. MBG analyzed the data and developed the theoretical model, assisted by BDG. MBG and BDG cowrote the paper with input from all authors.
**Competing interests:** Authors declare no competing interests.

# Supplementary Materials for

## Coulomb blockade in an atomically thin quantum dot coupled to a tunable Fermi reservoir

Mauro Brotons-Gisbert, Artur Branny, Santosh Kumar, Raphaël Picard, Raphaël Proux, Mason Gray, Kenneth S. Burch, Kenji Watanabe, Takashi Taniguchi, Brian D. Gerardot

Correspondence to: M.Brotons_i_Gisbert@hw.ac.uk or B.D.Gerardot@hw.ac.uk;

## Materials and Methods

### Device fabrication

Mechanical exfoliation from bulk crystals was used for each 2D material. The number of layers of $WSe_2$ (from HQ Graphene) and hBN flakes was initially determined by the optical contrast. The all-dry transfer technique[1] was utilized to construct the vdW heterostructure presented in the main text (Fig. 1a). Supplementary Figure 1a shows an optical image of the full device. The heterostructure assembly took place on a n-type Si wafer with a thermal oxide layer (from Graphene Supermarket). The thicknesses of the oxide layer and of the bottom hBN flake were measured using nulling ellipsometry (Accurion EP4) to be 98 nm and 7.3 nm, respectively. The ML thickness of the $WSe_2$ flake (Suppl. Fig. 1b) was confirmed by its PL emission energy. A thickness of $0.5 \pm 0.2$ nm was obtained for the hBN tunnel barrier from the combination of ellipsometry measurements and the lever arm of the device, which was obtained from the fit of the experimental dependence of our QD PL energy as function of the gate voltage to the zero-bandwidth model (see Fig. 2b). Supplementary Figure 1c shows a color-coded spatial maps of the integrated PL signal of the corresponding $WSe_2$ flake in the spectral range of 1.515 -1.722 eV (top panel) and 1.512 – 1.631 eV (bottom panel) highlighting the full flake and the single quantum emitters, respectively.

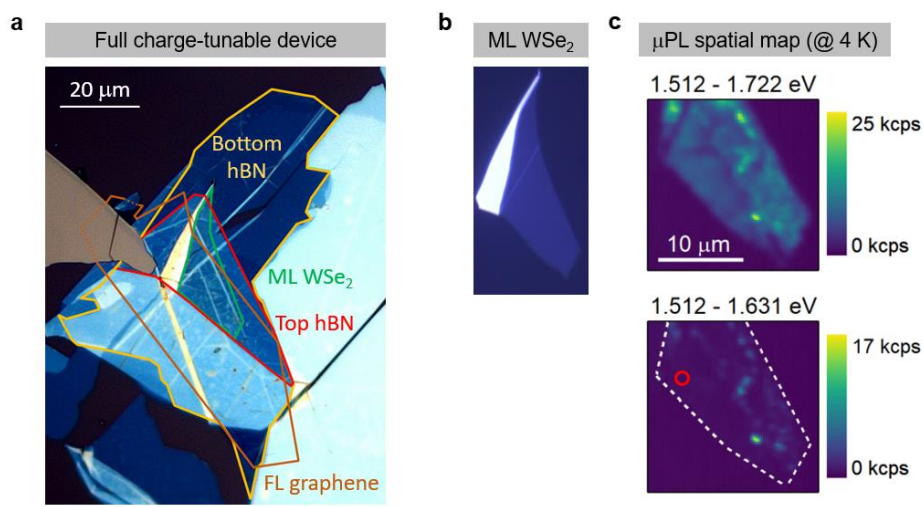

**Supplementary Figure 1. Charge-tunable vdW heterostructure. a**, Optical micrograph of completed device featuring a color-coded indication of the different layers employed in the heterostructure assembly. Note the wrinkles visible in the ML $WSe_2$ region (green outline). **b**, Micrograph of the ML $WSe_2$ flake before the transfer. **c**, Color-coded spatial maps of the integrated PL signal of the final device in the region of the $WSe_2$ flake (spectral range of 1.512 – 1.722 and 1.512 – 1.631 eV). These energy ranges highlight the full flake (top panel) and the quantum dots (bottom panel). The red circle represents a region of the sample where no emission from localized excitons was observed.

**Supplementary Text**

Charge tuning of the 2D exciton in ML WSe$_2$

To initially characterize the charge-tuning behaviour of the device, differential reflectivity measurements as a function of the applied bias were carried out in a region of the sample where no emission from localized excitons was observed (see red circle in Suppl. Fig. 1c). This experimental technique has been demonstrated to be a very useful technique to test both the sample quality and the charge-tuning characteristics of the 2D exciton in vdW heterostructures[2]. Our results (Suppl. Fig. 2) show that by changing the applied bias we are able to change from a neutral charge regime, in which the Fermi level lays between the valence and conduction bands of WSe$_2$, to a n-type regime in which the Fermi level lays above the conduction band of WSe$_2$. In the neutral regime we only observe the 2D neutral exciton transition 2D-$X^0$. However, in the n-doped regime we observe both the 2D-$X^0$ and the two transitions associated to the negatively charged exciton 2D-$X^{1-}$, which show a fine-structure splitting of ~7 meV and binding energies of ~33 and ~40 meV, in agreement with previous observations[2]. Supplementary Figure 2b shows reflected intensity spectra of ML WSe$_2$ for applied voltage gates of 0 (black solid line) and 18 V (red line), in which the resonances corresponding to the $X^{1-}$ and $X^0$ exciton states can be observed.

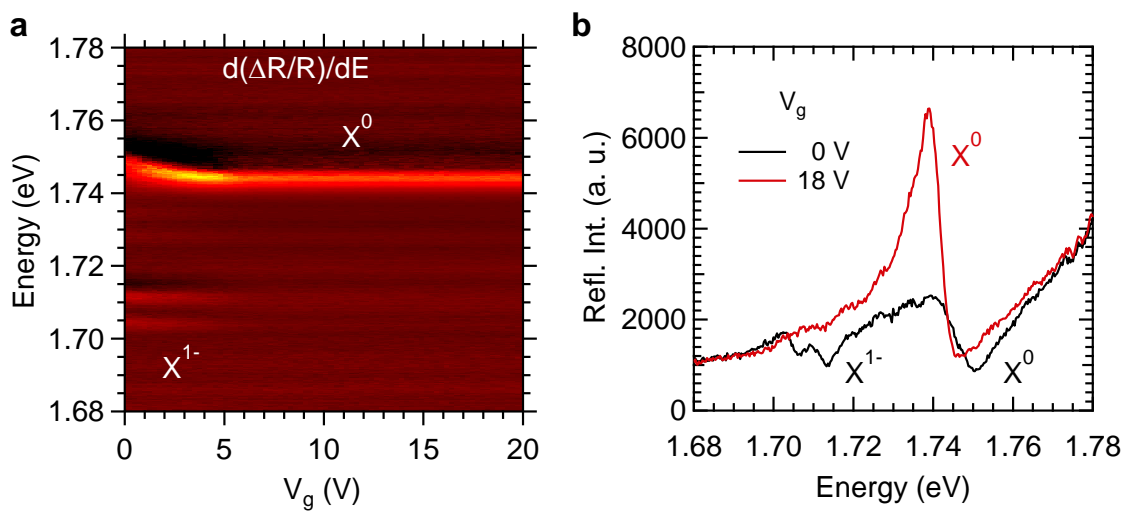

**Supplementary Figure 2. Charge tuning of the 2D exciton in ML WSe$_2$. a**, Contour plot of the first derivative with respect to energy of the differential reflectivity of the ML WSe$_2$ sample used in this work as a function of the applied bias. The measurements were carried out at the region of the sample identified in Fig. S1c (where no emission from localized excitons was observed). The $X^{1-}$ and $X^0$ charged states of the 2D exciton of WSe$_2$ are clearly observed. **b**, Reflected intensity spectra of ML WSe$_2$ for applied voltage gates of 0 (black solid line) and 18 V (red line), in which the resonances corresponding to the $X^{1-}$ and $X^0$ exciton states can be observed.

Additional features in Figure 1c in the main manuscript

In addition to the $X^{1-}$ and $X^0$ states of Dots A, B and C, the voltage-dependent PL shown in Fig. 1c in the main manuscript also reveals the presence of the $X^{1-}$ and $X^0$ states corresponding to two additional quantum dots (dots D and E) which also exhibit Coulomb blockade (Suppl. Fig. 3a). Dot D shows a PL energy and a voltage-dependent PL evolution very similar to Dot C, although it is possible to distinguish the emission of both quantum dots in higher spectral resolution measurements. Supplementary Figure 3b shows the voltage-dependent evolution of the PL of Dots C and D for applied voltages between 0.5 and 9.5 V measured with higher spectral resolution. As it can be seen in this figure, three emission lines can be clearly distinguished. Two of these lines (see labels in Suppl. Fig. 3b) correspond to the fine-structure-split emission doublet of the $X^0$ state of Dot C, which is the brightest emitter in the measurement. These two emission lines show the same exact energy evolution as a function of the applied voltage, since the fine-structure splitting remains constant for applied voltages far from the voltages at which hybridisation with the Fermi reservoir occurs. The third line corresponds to the low energy state of the fine-structure split $X^0$ doublet of Dot D. The very low emission intensity of dot D prevents the observation of the high energy state, which has an emission intensity below the noise level. To further prove that the three emission lines observed in Suppl. Fig. 3b correspond to the $X^0$ states of two different dots, Suppl. Fig. 3c shows the temporal evolution over ten minutes of the PL energy of Dots C and D measured at an applied voltage at which the emission peaks of dots C and D can be clearly resolved ($V_g = 9.5$ V). As it can be observed in this figure, the fine-structure-split peaks of the $X^0$ state of Dot C show identical spectral shifts over time due to spectral fluctuations, which further proves that the two emission lines belong to the same quantum dot. On the contrary, the temporal evolution of the PL energy of Dot D is uncorrelated to the emission peaks of Dot C, as expected for quantum dots which are at spatially different locations (and thus sensitive to different local charge environments).

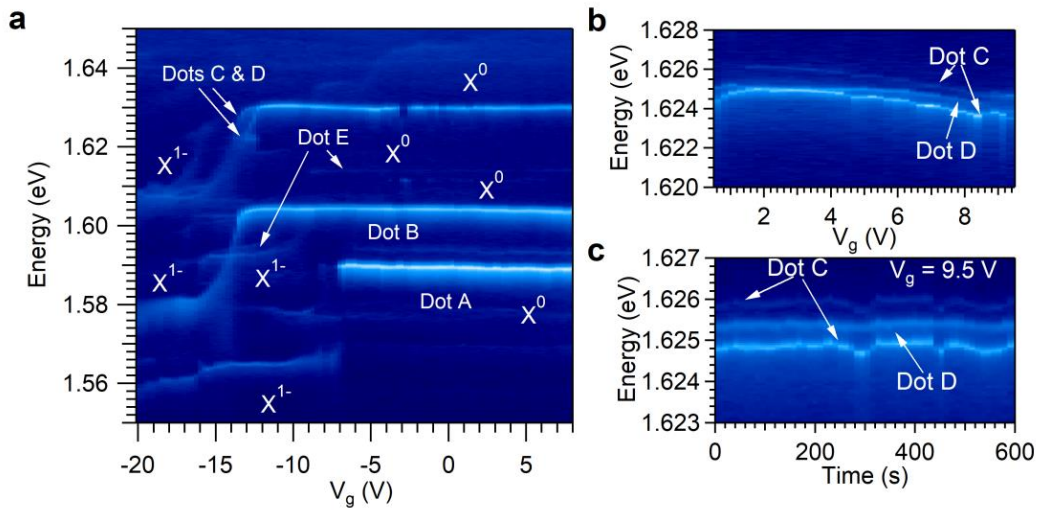

**Supplementary Figure 3. Additional features in Figure 1c in the main manuscript. a**, Voltage-dependent low-resolution PL of QDs A, B, C, D and E showing neutral ($X^0$) and negatively charged ($X^{1-}$) exciton species and Coulomb blockade. **b**, Voltage-dependent high-resolution PL of dots C and D for applied voltages between 0.5 and 9.5 V. **c**, Temporal evolution over ten minutes of the PL energy of dots C and D measured at an applied voltage of Vg = 9.5 V.

Polarization direction of the neutral exciton

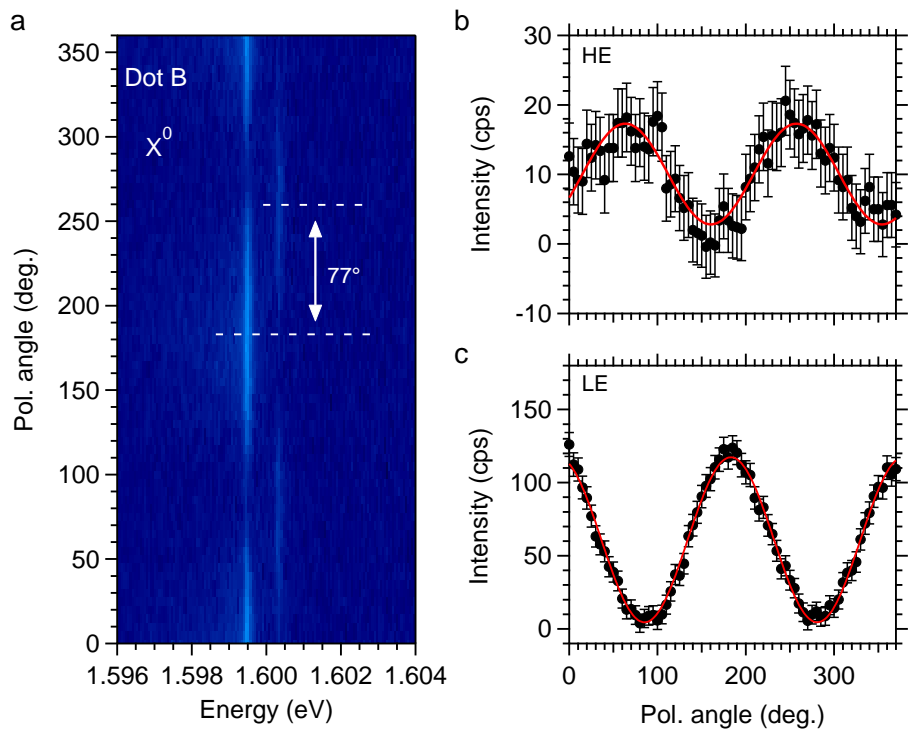

**Supplementary Figure 4. Polarization emission of the neutral exciton. a**, Color-coded PL intensity as a function of polarization for the neutral exciton of QD B measured at $V_g = 0$. The two dashed horizontal lines indicate the polarization direction of the low-energy (LE) and high-energy (HE) emission lines. **b, c,** Integrated intensities of the high-energy (**b**) and low-energy (**c**) emission lines, respectively, as a function of polarization rotation angle as extracted from **a**. The red solid lines represent fits of the experimental data to a cosine squared function.

Lineshape evolution of the exciton states as a function of the gate voltage

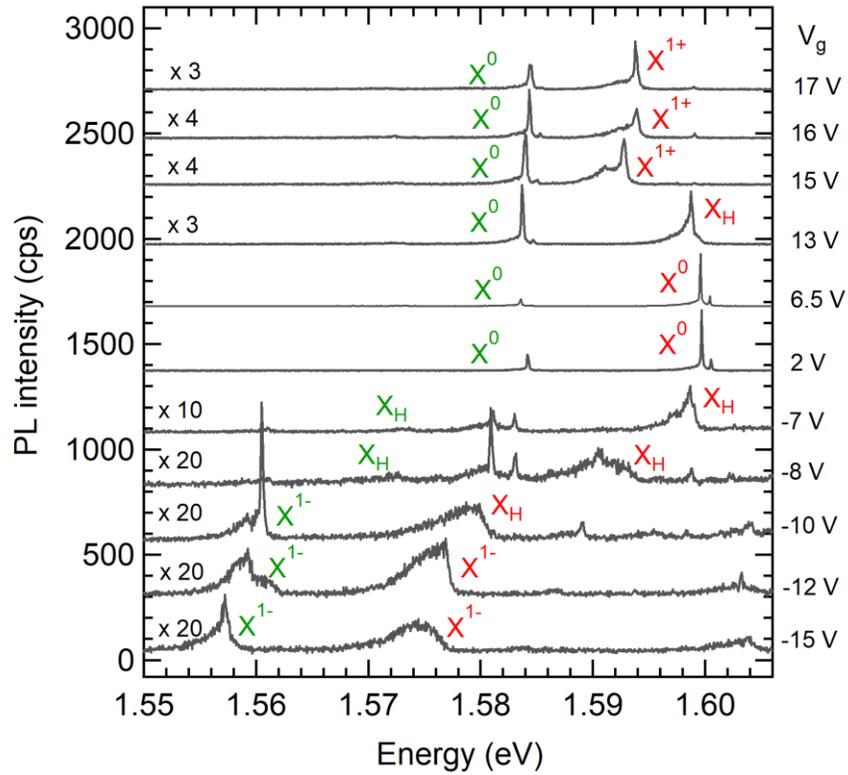

**Supplementary Figure 5. Lineshape evolution of the excitons states as a function of the gate voltage.** Photoluminescence spectra of quantum dots A and B for different applied gate voltage values measured at T = 3.8 K. The green and red labels indicate the charged exciton states for quantum dots A and B, respectively. Labels $X^{1-}$, $X^0$, $X^{1+}$ and $X_H$ represent the negatively charged, neutral, positively charged and hybrid excitons, respectively. For comparison purposes, some spectra have been multiplied by the corresponding factors indicated at the left part of the figure.

## Zero-bandwidth model with fine-structure splitting

### Zero magnetic field $(\vec{B}_z = 0)$

The zero-bandwidth model is a simplified model in which the Fermi sea is replaced by a single "delocalized" quantum level of energy $E_F$[3-5]. Although this model does not provide optical line shapes, it is a very useful model to explain the presence of different hybridized exciton states and their physical properties. Figure 2d (main text) shows diagrams of the initial $|i\rangle$ and final $|f\rangle$ states considered in our model for the hybridization of the $X^0$ and $X^{1-}$ exciton states. Since our aim is to explore the hybridization of the $X^0$ and $X^{1-}$ exciton states, the initial state contains two electrons and one hole (one electron and two holes for the hybridization of the $X^0$ and $X^{1+}$ excitons). After photon emission from the state $|i\rangle$, the state $|f\rangle$ contains only one electron and it can be expressed as a superposition of two states[4]

$$|f\rangle = A_1|f_1\rangle + A_2|f_2\rangle, \tag{S1}$$

with

$$|f_1\rangle = \frac{1}{\sqrt{2}}(|FR^\uparrow\rangle - |FR^\downarrow\rangle), \tag{S2}$$

and

$$|f_2\rangle = \frac{1}{\sqrt{2}}(|QD^\uparrow\rangle - |QD^\downarrow\rangle), \tag{S3}$$

where the superscript single arrows account for the possible orientations of the electron spin. The state $|f_1\rangle$ is a linear combination of the spin orientation states of the remaining electron in the Fermi reservoir ($|FR\rangle$), whereas the state $|f_2\rangle$ is a superposition of the spin orientations for the remaining electron in the quantum dot. $|QD^\uparrow\rangle$ and $|QD^\downarrow\rangle$ are degenerate in energy. In the final hybridized state expressed in Eq. S1, the states $|f_1\rangle$ and $|f_2\rangle$ are coupled by a spin-conserving tunnel coupling with strength $V_{tun}$ and the relative contributions of $A_1$ and $A_2$ depend on $V_g$ (Ref.[4]). The Hamiltonian corresponding to this final state configuration can thus be written in the basis $\{|f_1\rangle, |f_2\rangle\}$ as:

$$H_f = \begin{pmatrix} E_{FR} & V_{tun} \\ V_{tun} & E_S \end{pmatrix}, \tag{S4}$$

where $E_{FR} = E_F$ is the energy of the electron (hole) in the Fermi reservoir and $E_S$ is the energy of the s-shell in the QD, which can be related to the Fermi energy through the applied voltage $\Delta V_g$ and the ratio of the device thickness to tunnel barrier thickness (referred to as the lever arm $\lambda$ of the device)[4]:

$$E_S = -e\frac{\Delta V_g}{\lambda} + E_F, \tag{S5}$$

where $e$ is the magnitude of the electron charge. In our calculation, we may choose the zero of energy at our convenience, which we do by setting $E_F = 0$. By doing this, the QD energy is directly given by the applied voltage and the lever arm. Moreover, in order to take into account the voltage $V_c$ at which the first electron tunnels into the empty QD from the graphene contact, we re-express $\Delta V_g$ as $\Delta V_g = V_g - V_c$. By diagonalizing the Hamiltonian in Eq. S4, we find the hybridized final state energy:

$$E_{f,\pm} = \frac{E_{FR} + E_S \pm \sqrt{(E_{FR} - E_S)^2 + 4V_{tun}^2}}{2}. \tag{S6}$$

The initial excitonic state is also a superposition of different excitonic states (see Fig. 2d in the main text):

$$|i\rangle = \frac{B_1}{\sqrt{2}}(|i_1\rangle - |i_2\rangle) + \frac{B_2}{\sqrt{2}}(|i_3\rangle - |i_4\rangle), \tag{S7}$$

with

$$|i_1\rangle = \frac{1}{\sqrt{2}}\left(|QD^{\Uparrow\downarrow}, FR^{\uparrow}\rangle + |QD^{\Downarrow\uparrow}, FR^{\downarrow}\rangle\right), \tag{S8}$$

$$|i_2\rangle = \frac{1}{\sqrt{2}}\left(|QD^{\Uparrow\downarrow}, FR^{\uparrow}\rangle - |QD^{\Downarrow\uparrow}, FR^{\downarrow}\rangle\right), \tag{S9}$$

$$|i_3\rangle = |QD^{\Uparrow\downarrow\uparrow}\rangle, \text{ and} \tag{S10}$$

$$|i_4\rangle = |QD^{\Downarrow\uparrow\downarrow}\rangle. \tag{S11}$$

The states $|i_1\rangle$ and $|i_2\rangle$ correspond to the states in which the QD contains one electron-hole pair (giving rise to the neutral exciton $X^0$) and an additional electron is in the Fermi reservoir. The superscript single (double) arrows account for the possible orientations of the electron (hole) spin. $B_1$ and $B_2$ are amplitudes which can be continuously tuned by $V_g$. Since in $|i_1\rangle$ and $|i_2\rangle$ the QD only contains an electron-hole pair, the exchange interaction splits the $X^0$ into two states with energies $E_{X^0} + \Delta_{FSS}/2$ ($|i_1\rangle$) and $E_{X^0} - \Delta_{FSS}/2$ ($|i_2\rangle$), where $E_{X^0}$ is the central energy of the emission doublet and $\Delta_{FSS}$ is the so called fine-structure splitting energy. The states $|i_3\rangle$ and $|i_4\rangle$ correspond to the excitonic configurations in which the QD has two electrons with opposite spin orientations and a hole, giving rise to $X^{1-}$ with an energy $E_{X^-}$. This excitonic state can be considered as a hole interacting with a spin singlet electron pair and therefore, the electron-hole exchange energy splitting vanishes[6]. The individual states in $|i\rangle$ are coupled by a spin-conserving tunnel interaction with strength $V_{tun}$ between the electrons in the QD and the Fermi reservoir: $|QD^{\Uparrow\downarrow}, FR^{\uparrow}\rangle$ couples to $|QD^{\Uparrow\downarrow\uparrow}\rangle$, and $|QD^{\Downarrow\uparrow}, FR^{\downarrow}\rangle$ couples to $|QD^{\Downarrow\uparrow\downarrow}\rangle$. The Hamiltonian corresponding to this initial state configuration can be expressed in the basis $\{|QD^{\Uparrow\downarrow}, FR^{\uparrow}\rangle, |QD^{\Downarrow\uparrow}, FR^{\downarrow}\rangle, |QD^{\Uparrow\downarrow\uparrow}\rangle, |QD^{\Downarrow\uparrow\downarrow}\rangle\}$ as follows:

$$H_i = \begin{pmatrix} E_{X^0} & \Delta_{FSS}/2 & \sqrt{2}V_{tun} & 0 \\ \Delta_{FSS}/2 & E_{X^0} & 0 & \sqrt{2}V_{tun} \\ \sqrt{2}V_{tun} & 0 & E_{X^-} & 0 \\ 0 & \sqrt{2}V_{tun} & 0 & E_{X^-} \end{pmatrix}. \tag{S12}$$

The energies of the hybridized initial state can be obtained by diagonalizing the Hamiltonian in Eq. S12. At low temperatures, we take only the initial states with the lowest energies:

$$E_i^{\pm} = \frac{E_{X^0} \pm \Delta_{FSS}/2 + E_{X^-} - \sqrt{(E_{X^0} \pm \Delta_{FSS}/2 - E_{X^-})^2 + 8V_{tun}^2}}{2}, \tag{S13}$$

where $E_{X^0} = \hbar\omega_{X^0}$ and $E_{X^{1-}} = \hbar\omega_{X^{1-}} - e\,\Delta V_g/\lambda$, with $\hbar\omega_{X^0}$ and $\hbar\omega_{X^{1-}}$ being the energies observed for the neutral and negatively charged excitons, respectively.

The energies $E_{X_H}$ of the resultant hybrid excitons can then be calculated as

$$E_{X_H,upper(lower)}^{\pm}(\Delta V_g) = E_i^{\pm} - E_{f,\mp}. \tag{S14}$$

As it can be seen in the previous equation, the emission spectrum in the zero-bandwidth model has several emission lines[3,4]. However, the most intense lines at low temperature are

$$E_{X_H,upper}^{\pm}(\Delta V_g) = E_i^{\pm} - E_{f,-}. \tag{S15}$$

Similar to the hybridization of the $X^0$ and $X^{1-}$ exciton states, the zero-bandwidth model can be used to explore the hybridization of the $X^0$ and $X^{1+}$ exciton states as well. In this case, the initial state contains one electron and two holes, and after photon emission from the initial state the final state $|f\rangle$ only contains one hole. It is straightforward then to realize that the hybridization between the $X^0$ and $X^{1-}$ and between the $X^0$ and $X^{1+}$ exciton states can be modelled by using exactly the same set of equations due the analogy of the initial and final states for both cases.

Supplementary Figure 6a shows the calculated evolution of the energies of the two lowest hybrid initial states (upper panel) and the energies of the two hybrid final states (lower panel) as a function of $V_g$ for the $X^0$ - $X^{1+}$ transition. For the calculation, we have employed the same set of parameters described in the main text to fit the experimental results for the $X^0$ - $X^{1+}$ transition. Supplementary Figure 6b shows the calculated evolution with $V_g$ of the hybrid excitons resulting from the initial and final states shown in Suppl. Fig. 6a. The color and line style of the solutions shown in this figure are consistent with the ones employed in Suppl. Fig. 6a. In this way, red and blue solid lines in Suppl. Fig. 6b represent $E_{X_H,upper}^+$ and $E_{X_H,upper}^-$, respectively, which originate from transitions between the initial states $E_i^+$ and $E_i^-$, and the final state $E_{f,-}$. As mentioned before, these lines are the most intense ones at low temperature.

The hybrid excitons originating from transitions involving the initial states $E_i^+$ and $E_i^-$ and the final state $E_{f,+}$ are represented in Suppl. Fig. 6b by red and blue dashed lines, respectively. This set of solutions, labelled as $X_{f+}$, corresponds to the situation in which the remaining particle in the final state is tunneling out of the QD, as discussed in Refs.[4,7]. As can be observed, the set of solutions $X_{f+}$ are split from the upper set of solutions. Such splitting has its origin in the anti-crossing of the final states (see lower panel of Suppl. Fig. 6a), a consequence of the zero-bandwidth character of the model[4]. On the contrary, if the Fermi reservoir is modeled as a continuum of states rather than a single delocalized model, a continuous transition from the $X_{f+}$ to the $X^{1+}$ is obtained (see Suppl. Fig. 8).

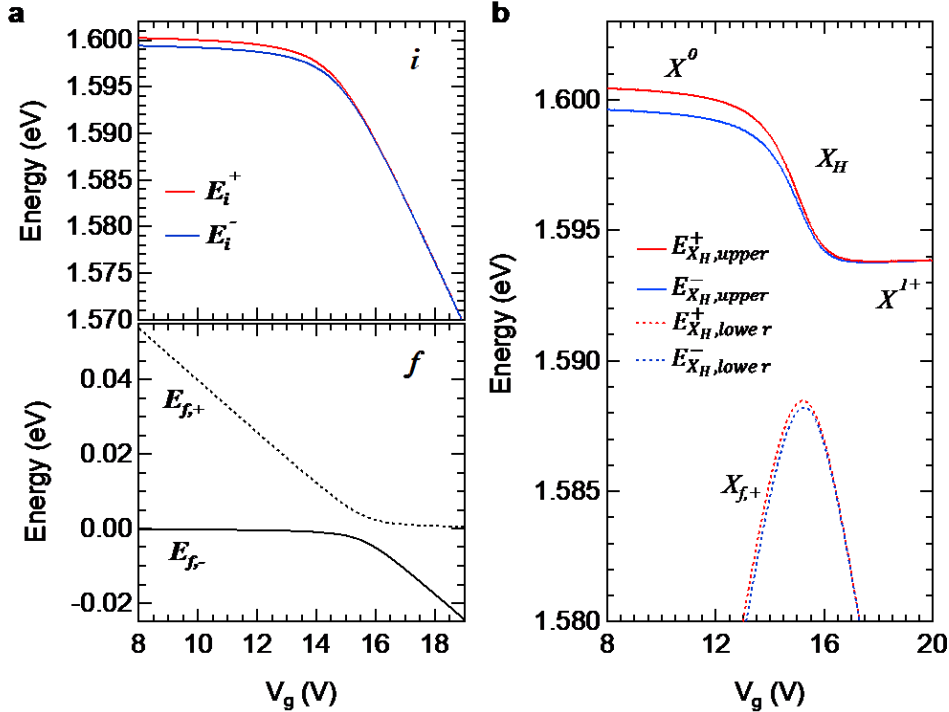

**Supplementary Figure 6. Zero-bandwidth model with fine-structure splitting. a**, Calculated evolution of the energies of the two lowest hybrid initial states (upper panel) and the energies of the two hybrid final states (lower panel) as a function of the applied gate voltage for the $X^0$ - $X^{1+}$ transition. For the calculation, we have employed the same set of parameters described in the main text to fit the experimental results for the $X^0$ - $X^{1+}$ transition. **b**, Calculated evolution of the hybrid excitons resulting from the initial and final states shown in panel **a** as a function of the applied gate voltage. The color and line style of the solutions shown in this figure are consistent with the ones employed in part **a**.

**Non-zero magnetic field in Faraday geometry $\left(\vec{B}_z \neq 0\right)$**

We also explore theoretically the hybridization between the $X^0$ and $X^{1-}$ and between the $X^0$ and $X^{1+}$ exciton states when a magnetic field $\vec{B}_z$ is applied along the direction perpendicular to heterostructure interfaces (Faraday geometry). The zero-bandwidth model previously described can be adapted by including the effects that the applied magnetic field has on both the hybridized initial and final states. It is worth noticing that due to the applied magnetic field, the states $|QD^\uparrow\rangle$ and $|QD^\downarrow\rangle$ referring to an electron (or hole) remaining in the QD after photon emission are no longer degenerate in energy. For this reason, the Hamiltonian corresponding to the hybridized final state under a non-zero magnetic field in Faraday geometry can be written as a 4x4 matrix:

$$H_{f,B_z}(B_z) = \begin{pmatrix} E_{FR} & V_{tun} & 0 & 0 \\ V_{tun} & E_s + \Delta_{B,e(h)} & 0 & 0 \\ 0 & 0 & E_{FR} & V_{tun} \\ 0 & 0 & V_{tun} & E_s - \Delta_{B,e(h)} \end{pmatrix}, \quad (S16)$$

where $\Delta_{B,e(h)} = \mu_B g_{e(h)} B_z / 2$, with $\mu_B$ being the Bohr magneton and $g_{e(h)}$ being the gyromagnetic ratio for the remaining electron (hole). It is important to notice that the expression for $H_{f,B_z}(B_z)$ is written under the assumption that the single "delocalized" quantum level that replaces the Fermi reservoir $|FR\rangle$ remains degenerate in energy for both opposite directions of the electron (hole) spin. The energies of the hybridized final state can be obtained by diagonalizing the Hamiltonian in Eq. (S16). Again, at low temperatures we take only the final states with the lowest energies:

$$E_{f,B_z}^\pm = \frac{E_{FR} + E_s \pm \Delta_{B,e(h)} - \sqrt{\left(E_{FR} - (E_s \pm \Delta_{B,e(h)})\right)^2 + 4V_{tun}^2}}{2}. \quad (S17)$$

As commented before, it is also necessary to include the effects that the applied magnetic field has on the hybridized initial states. On the one hand, under $\vec{B}_z$ the energy splitting of the $X^0$ fine structure doublet increases by the Zeeman interaction of the electron and hole spins. On the other hand, $\vec{B}_z$ also causes an energy splitting of the $X^{1\pm}$ trions, thus breaking the energy degeneracy of the states $|QD^{\Uparrow\downarrow\uparrow}\rangle$ and $|QD^{\Downarrow\uparrow\downarrow}\rangle$. All these effects can be included in the Hamiltonian of the initial state corresponding to the zero-magnetic field case (Eq. S12). In this way, the Hamiltonian corresponding to the initial state configuration under a non-zero magnetic field can be expressed in the basis $\{|QD^{\Uparrow\downarrow}, FR^\uparrow\rangle, |QD^{\Downarrow\uparrow}, FR^\downarrow\rangle, |QD^{\Uparrow\downarrow\uparrow}\rangle, |QD^{\Downarrow\uparrow\downarrow}\rangle\}$ as follows:

$$E_{i,B_z}^\pm = \begin{pmatrix} E_0 & \Delta_{B,X^0} & \sqrt{2}V_{tun} & 0 \\ \Delta_{B,X^0} & E_0 & 0 & \sqrt{2}V_{tun} \\ \sqrt{2}V_{tun} & 0 & E_{X^-} & \Delta_{B,h(e)} \\ 0 & \sqrt{2}V_{tun} & \Delta_{B,h(e)} & E_{X^-} \end{pmatrix}, \quad (S18)$$

where $\Delta_{B,X^0} = \sqrt{\Delta_0^2 + (\mu_B(g_e + g_h)B_z)^2}/2$ and $\Delta_{B,h(e)} = (\mu_B g_{h(e)} B_z)/2$, with $g_{h(e)}$ the gyromagnetic ratio for the hole (electron). The energies of the hybridized initial state can be obtained by diagonalizing the Hamiltonian in Eq. S18. Again, at low temperatures we take only the initial states with the lowest energies:

$$E_{i,B_z}^\pm = \frac{E_{X^0} \pm \Delta_{B,X^0} + E_{X^-} \pm \Delta_{B,h(e)} - \sqrt{\left(E_{X^0} \pm \Delta_{B,X^0} - (E_{X^-} \pm \Delta_{B,h(e)})\right)^2 + 8V_{tun}^2}}{2}. \quad (S19)$$

Supplementary Figure 7a shows the calculated evolution of the energies of the two lowest hybrid initial states (upper panel) and the energies of the four hybrid final states (lower panel) as a function of the applied gate voltage for the $X^0$ - $X^{1+}$ transition as predicted by the zero-bandwidth model for $\vec{B}_z = 5.3$ T. For the calculation, we have employed the same set of parameters used to fit experimental results in Fig. 3b for the $X^0$ - $X^{1+}$ transition. Supplementary Figure 7b shows the calculated evolution with $V_g$ of the hybrid excitons resulting from the initial and final states shown in Suppl. Fig. 7a. The color and line style of the solutions shown in this figure are consistent with the ones employed in Suppl. Fig. 7a. As an example, the red line in Suppl. Fig. 7b represents the hybrid exciton resulting from the initial state $E_i^+$ and the final state represented by a red solid line. The blue dashed line in Suppl. Fig. 7b represents a hybrid exciton originating from the initial state $E_i^-$ (black dashed line) and the final state represented by a blue solid line.

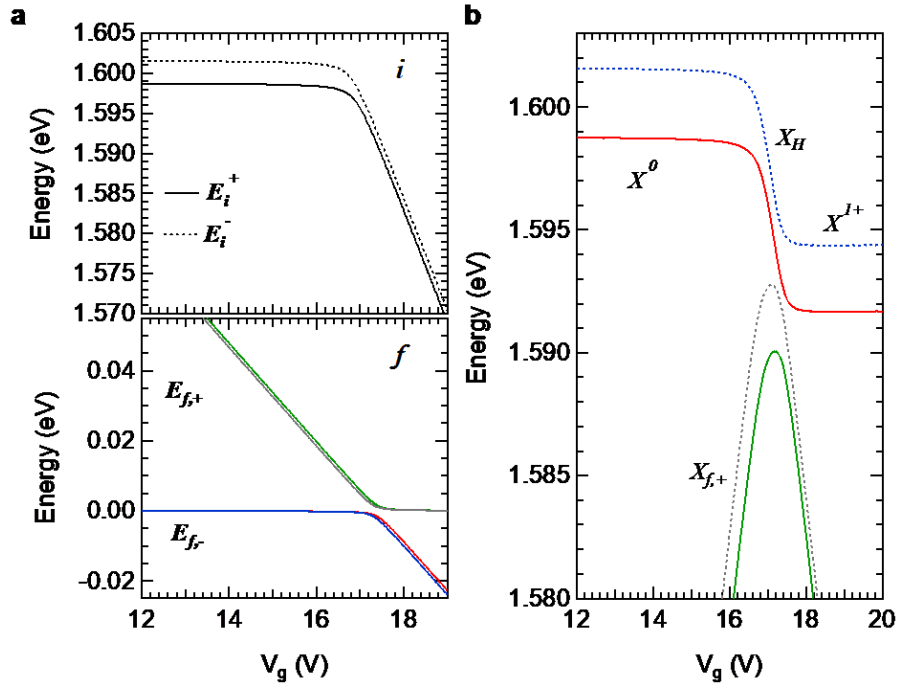

**Supplementary Figure 7. Zero-bandwidth model with fine-structure splitting under a non-zero magnetic field. a**, Calculated evolution of the energies of the two lowest hybrid initial states (upper panel) and the energies of the four hybrid final states (lower panel) as a function of the applied gate voltage for the $X^0$ - $X^{1+}$ transition as predicted by the zero-bandwidth model for $\vec{B}_z = 5.3$ T. For the calculation, we have employed the same set of parameters described in the main text to fit the experimental results for the $X^0$ - $X^{1+}$ transition shown in Fig. 3b in the main text. **b**, Calculated evolution of the hybrid excitons resulting from the initial and final states shown in **a** as a function of the gate voltage. The color and line style of the solutions shown in this figure are consistent with the ones employed in **a**.

$X_f$ - $X^{1+}$ transition

The zero-bandwidth model predicts the existence of the $X_{f+}$ feature observed in the experimental results (see Suppl. Figs. 6b and 7b), although it introduces an artificial splitting between the $X_{f+}$ and $X^{1+}$ plateau due to the simplicity of the model. On the contrary, if the Fermi reservoir is modelled as a continuum of states rather than a single delocalized level, a continuous transition from the $X_{f+}$ to the $X^{1+}$ is obtained[7-9]. The blue dots in Suppl. Fig. 8 represent the measured PL energy as a function of the applied $V_g$ for the $X^{1+}$ exciton state of QD B under an applied magnetic field of 570 mT. The colour code in this figure represents the calculated PL intensity of the $X^{1+}$ state for a QD strongly coupled to a Fermi reservoir using the analytical model presented in Ref.[7]. This model assumes that the system involving a QD strongly coupled to a Fermi sea can be described by the Anderson Hamiltonian and that low energy excitations of the Fermi sea are unimportant, essentially treating the Fermi sea as a frozen spectator to the tunneling processes. In the calculation, we considered a T = 3.8 K and the lever arm and tunnel coupling energy obtained from the experimental results. The best agreement between the experimental and calculated evolution of the PL energy has been obtained with $U_{hh} = 26$ meV, a bandwidth of 30 meV for the occupied states at the hole reservoir (flat DOS), and a Stark shift of ~200 μeV/V. While this simplistic model describes the experimental data, a more complete model using Wilson's numerical renormalization group method can be used to explain the many-body interactions[7, 10, 11].

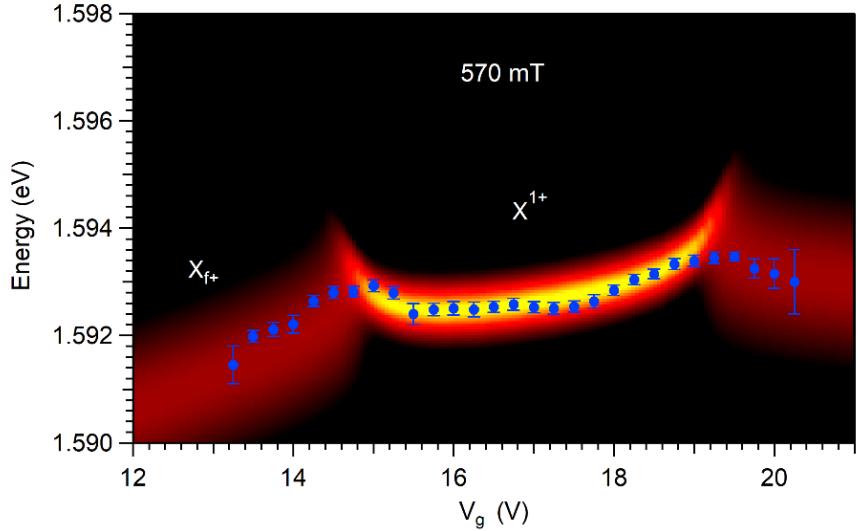

**Supplementary Figure 8. $X_{f+}$ feature.** Measured PL energy as a function of the applied $V_g$ for the $X^{1+}$ exciton state of QD B under an applied magnetic field of 570 mT (blue dots). The colour code in this figure represents the calculated PL intensity of the $X^{1+}$ state for a QD strongly coupled to a Fermi reservoir.

Zero-bandwidth and Coulomb blockade model for Dots A, C and E

Dots A - E all exhibit Coulomb blockade and can be explained with the zero-bandwidth Anderson impurity model. Supplementary Figure 9 below shows the fits to the experimental data for dots A, C and E, where the solid lines are solutions to the analytical model. Here we extract binding energies of ~ 24.4, ~ 22.4, and ~ 21.3 meV and tunnel coupling energies of ~0, 3.2 ± 0.1, and 2.6 ± 1.2 meV for Dots A, C, and E, respectively. We expect the tunnel coupling energy for Dot A to be finite, but its value is less than what we can resolve experimentally.

Additionally, according to the Coulomb blockade model, for the $X^0 \leftrightarrow X^{1-}$ transition the electron tunnels into the QD when it can overcome the electron-electron Coulomb interaction. This charging energy should depend on the QD size and thus on its emission energy. So, for a collection of QDs in the same device (and therefore same lever arm), one expects to see the onset of the tunnelling events (or charging thresholds) to be dependent on emission energy. In general, we observe such a trend in the charging thresholds, as can be seen in Suppl. Fig. 3a above as well as in Suppl. Fig. 9 below.

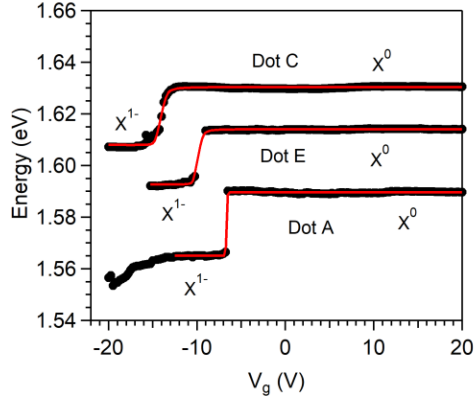

**Supplementary Figure 9. Zero-bandwidth model for dots A, C and E.** Voltage-dependent evolution of the emission energy for the QDs A, C and E extracted from Suppl. Fig. 3a, showing Coulomb blockade. The solid lines represent fits of the experimental data using the zero-bandwidth Anderson impurity model.

Electron-hole tunnel coupling ratio

As discussed in the main text, a tunnel-coupling ratio $V_{tun}^h/V_{tun}^e = 2.9 \pm 0.5$ can be directly extracted from the fits shown in Fig. 2b. This ratio is a consequence of the band alignment resulting from the ML WSe$_2$/hBN/graphene heterostructure, which leads to significantly lower tunnel barrier heights for holes than for electrons[12]. $V_{tun}^h/V_{tun}^e$ can be estimated using the Wentzel-Kramer-Brillouin (WKB) approximation

$$\frac{V_{tun}^h}{V_{tun}^e} = \frac{m_e^*}{m_h^*} \exp\left[2L\left(\sqrt{\frac{2m_e^*}{\hbar^2}\Delta V_e} - \sqrt{\frac{2m_h^*}{\hbar^2}\Delta V_h}\right)\right], \tag{S20}$$

where $L$ is the thickness of a rectangular tunneling barrier, $m_{e(h)}^*$ are the effective masses for the electron (holes), and $\Delta V_{e(h)}$ are the respective band offsets. Using $m_e^* = 0.34 m_0$ (Ref. [12]), $m_h^* = 0.45 m_0$ (Ref. [13]), $L$ is the thickness of an hBN ML, and tunnel barrier heights of $\Delta V_e \sim 3.3 - 3.5$ eV and $\Delta V_h \sim 0.9 - 1.1$ eV estimated from previous reports[12], we estimate a ratio $V_{tun}^h/V_{tun}^e \sim 3.1 \pm 0.4$, in agreement with the experimental result.

Tunnel-induced broadening of the emission linewidth

The WKB approximation can also be employed to estimate the tunnel-induced broadening of the PL for Dot B of as a function of $V_g$, as shown in the bottom panel of Fig. 2c. The tunneling process for electrons and holes between the QD and the Fermi sea leads to a reduced lifetime of the exciton in the dot and thus leads to an increase of the emission linewidth $\Gamma$. We estimate the tunnel-induced broadening of the emission linewidth by employing a semiclassical model[14]. In this model, and within the WKB approximation, the decay probability ($\tau^{-1}$) can be calculated as the transmission coefficient through a rectangular potential with a lateral extension $d$ and depth $\Delta V$ multiplied by the frequency of collisions with the wall. The frequency of wall collisions is connected with the velocity $v$ of the particle, which is estimated via the uncertainty principle as $v \gg \hbar/(2m_{e(h)}^* d)$. From the decay time $\tau$ it is possible to estimate the contribution of this tunneling process to the emission linewidth by means of the uncertainty relation $\Gamma \tau > \hbar/2$. This leads to the following expression:

$$\Gamma \geq \Gamma_0 + \frac{\hbar^2}{8m_{e(h)}^* d^2} T(\Delta V_g), \tag{S21}$$

where $\Gamma_0$ is the emission linewidth corresponding to the pure $X^0$ QD-state, $m^*_{e(h)}$ are the effective masses for the electron or holes, and $T(\Delta V_g)$ is the $V_g$-dependent transmission coefficient of a rectangular barrier for a non-resonant tunnelling[15]. Figure 2d (lower panel) in the main text shows a comparison between the experimental and calculated values for FWHM of the emission linewidth $\Gamma$. In the calculations, we have assumed a tunnel barrier thickness corresponding to a single h-BN layer, a lateral extension $d$ corresponding to a ML WSe$_2$ for the rectangular potential, and the carrier effective masses and band offsets described in the previous section.

Magneto-optics

Supplementary Figure 10 shows the voltage-dependent high-resolution PL of the $X^{1-}$ (left panels), $X^0$ (central panels) and $X^{1+}$ (right panels) exciton sates of QD B under different applied magnetic fields in Faraday geometry.

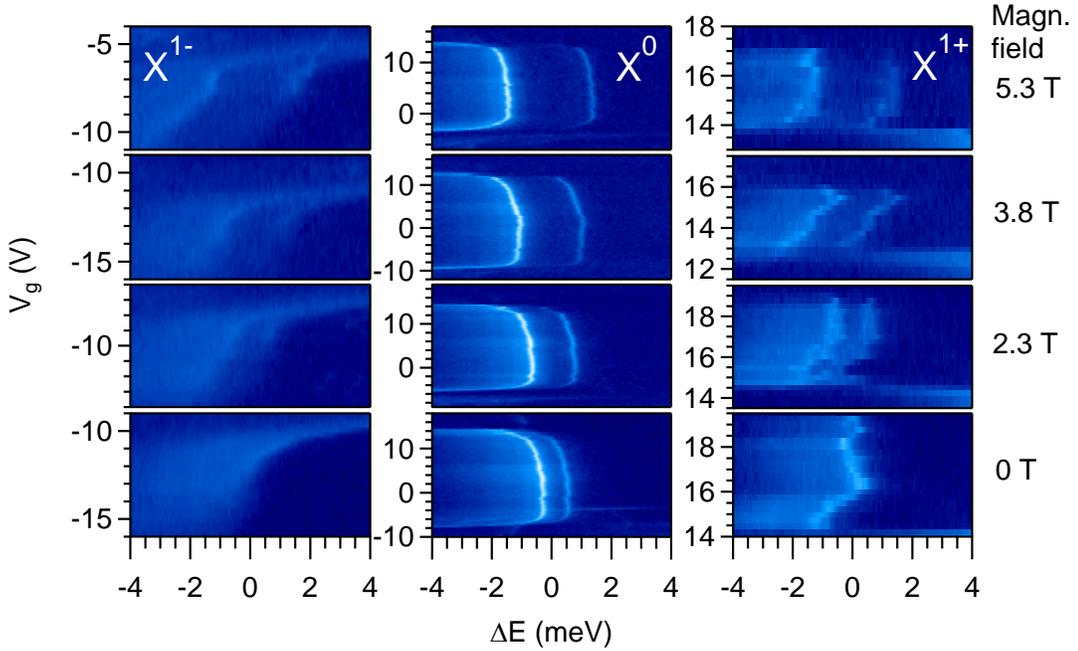

**Supplementary Figure 10. Voltage-dependent magneto-optics of different charged states in a quantum dot in ML WSe$_2$.** Voltage-dependent high-resolution PL of the $X^{1-}$ (left panels), $X^0$ (central panels) and $X^{1+}$ (right panels) exciton sates of QD B under different applied magnetic fields in Faraday geometry.

Finally, Figs. 4b and 4c in the main text show the magnetic-field dependence of the energy splitting measured for the $X^0$ doublet and the $X^{1-}$ and $X^{1+}$ excitons of Dot B, and for the $X^0$ doublet and the $X^{1-}$ exciton states of Dot A, respectively, as obtained from fits of the experimental data. Solid lines in Figs. 4b and 4c represent fits of the experimental data to equations

$$\Delta_{X^0} = \sqrt{\Delta_0^2 + (\mu_B g_{X^0} B_z)^2}, \quad \text{(S22)}$$

and

$$\Delta_{X^{1\pm}} = (\mu_B g_{X^{1\pm}} B_z), \quad \text{(S23)}$$

with $\mu_B$ being the Bohr magneton and $g_{X^0}$ and $g_{X^{1\pm}}$ being the gyromagnetic ratio for the neutral and charged excitons, respectively.